# Magnetization Governed Magnetoresistance Anisotropy in Topological Semimetal CeBi

Yang-Yang Lyu[1,2], Fei Han[3], Zhi-Li Xiao[1,4,*], Jing Xu[5], Yong-Lei Wang[2,*], Hua-Bing Wang[2], Jin-Ke Bao[1], Duck Young Chung[1], Mingda Li[3,*], Ivar Martin[1], Ulrich Welp[1], Mercouri G. Kanatzidis[1,6], and Wai-Kwong Kwok[1]

[1]*Materials Science Division, Argonne National Laboratory, Argonne, Illinois 60439, USA*

[2]*Research Institute of Superconductor Electronics, School of Electronic Science and Engineering, Nanjing University, Nanjing 210093, China*

[3]*Department of Nuclear Science and Engineering, Massachusetts Institute of Technology, Cambridge, Massachusetts 02139, USA*

[4]*Department of Physics, Northern Illinois University, DeKalb, Illinois 60115, USA*

[5]*Center for Nanoscale Materials, Argonne National Laboratory, Argonne, Illinois 60439, USA*

[6]*Department of Chemistry, Northwestern University, Evanston, Illinois 60208, USA*

*Correspondence to: xiao@anl.gov; yongleiwang@nju.edu.cn; mingda@mit.edu

Magnetic topological semimetals, the latest member of topological quantum materials, are attracting extensive attention as they may lead to topologically-driven spintronics. Currently, magnetotransport investigations on these materials are focused on the anomalous Hall effect. Here, we report on the magnetoresistance anisotropy of topological semimetal CeBi, which has tunable magnetic structures arising from localized Ce $4f$ electrons and exhibits both negative and positive magnetoresistances, depending on the temperature. We found that the angle dependence of the negative magnetoresistance, regardless of its large variation with the magnitude of the magnetic field and with temperature, is solely dictated by the field-induced magnetization that is orientated along a primary crystalline axis and flops under the influence of a rotating magnetic field. The results reveal the strong interaction between conduction electrons and magnetization in CeBi. They also indicate that magnetoresistance anisotropy can be used to uncover the magnetic behavior and the correlation between transport phenomena and magnetism in magnetic topological semimetals.

Stimulated by the fascinating properties discovered in topological insulators [1], topological quantum materials have become an exciting frontier in condensed matter physics and materials science [2-8]. Among them, magnetic topological insulators with strong correlations between magnetism and nontrivial band topology exhibit exotic phenomena such as quantum anomalous Hall effect [5,8-11] and axion insulator state [12]. However, those novel properties were observed in thin films of magnetic topological insulators converted from known topological insulators by doping magnetic atoms, e.g., chromium-doped $(Bi,Sb)_2Te_3$ [10,11] or in a ferromagnet-topological insulator-ferromagnet (FM-TI-FM) sandwich heterostructure [12]. Their fabrication requires advanced molecular beam epitaxy techniques [10-12]. The random magnetic dopants also inevitably introduce disorder that can hinder further exploration of topological quantum effects in the material [13,14]. Thus, topological insulators with intrinsic magnetic ordering have been extensively sought, leading to the recent revelation of magnetic topological insulator $MnBi_2Te_4$ [13-20]. Since 2016, researchers have also been searching for other types of topological materials with intrinsic magnetic order and uncovered a number of magnetic topological semimetals [21-38], which may lead to topologically-driven spintronics [37] and have become a focus center in the field [13-38]. Extensive exploration of these materials has led to the discovery of exotic properties. For example, both axion insulator and Chern insulator states have been demonstrated in magnetic topological insulator $MnBi_2Te_4$ [20]. Magnetization driven giant nematic energy shift [27] and large intrinsic anomalous Hall effects were observed in magnetic topological semimetals [26,29-34]. A new phenomenon called singular angular magnetoresistance is discovered in magnetic Weyl semimetal CeAlGe system very recently [39].

Here we report on an intriguing magnetoresistance (MR) anisotropy in the magnetic topological semimetal CeBi, which reveals the unusually strong interaction between conduction electrons and magnetization. Below the Néel temperature $T_N$, we observed a large variation in the angle dependence of the MR with the magnitude of the rotating magnetic field $\mathbf{H_R}$ and the temperature. We found that the anisotropy of the negative MR, although seemingly intricate, can be wholly accounted for if the MR is solely determined by the magnetization that aligns along one of the primary crystalline axes and flops between them in a constant $\mathbf{H_R}$. These assumptions are validated by the scaling behavior of the magnetic field dependence of the resistance $R(H)$ obtained at various field orientations. They are further supported by the consistency of the experimental MR anisotropies with those derived from the $R(H)$ curve obtained with $\mathbf{H}$ along the primary crystalline axis by converting the $H$ into $\varphi$ using $\varphi = \arccos(H/H_R)$, with $\varphi \leq 45°$ being the angle between the applied magnetic field and the primary crystalline axis. Our results highlight the strong correlation between transport phenomena and magnetism in magnetic topological semimetals. This work also provides a new way of using MR anisotropy to uncover magnetic behaviors such as flops of the magnetic moment in both ferromagnetic and non-ferromagnetic states.

Following the first report of uncommon magnetic behavior in 1965 [40], CeBi was intensively investigated in the '70s-'80s [41-46]. Magnetization measurements revealed rich magnetic phases including antiferromagnetic, ferrimagnetic and ferromagnetic phases, depending on temperature and magnitude of the magnetic field [41,45]. In the context of topological materials, CeBi recently has attracted renewed interest as it has been identified to be a topological semimetal in angle-resolved photoemission spectroscopy (ARPES) experiments [38,47]. It is also the only known magnetic topological semimetal that shows extremely large magnetoresistance (XMR) [47]. Although magnetotransport measurements were conducted more than 20 years ago and anomalous Hall effects were also observed [48], MR investigations have been only reported in the form of temperature dependence of the resistance $R(T)$ at low magnetic fields (up to 3.7 T) [48] or at a high magnetic field of 9 T [47]. Furthermore, the magnetic field in both cases was applied along a primary crystalline axis. We present below detailed studies on the MR in CeBi, with emphasis on its dependences on the magnitude and orientation of the magnetic field.

We studied two crystals (Samples A and B) in a standard four-probe transport configuration and one bare crystal (Sample C) for magnetization measurements. Additional experimental details are presented in the supplement. The $R(T)$ curve for Sample A in zero field is presented in Fig.1a. The expected [41,48] paramagnetic to antiferromagnetic (AFM) transition at $T_N \approx 25$ K and the transition from type I AFM to type IA AFM at $T_N/2$ ($\approx 12.8$ K), upon cooling, can be clearly identified (see Fig.S1b for schematics on magnetic structures). Figure 1b presents $R(H)$ curves obtained at various temperatures with $\mathbf{H} \parallel \mathbf{c}$. Above $T_N$, MR decreases monotonically with increasing magnetic field whereas below $T_N/2$, it increases with field, resulting in an XMR of ~$10^4$ % at $T = 3$ K and $H = 9$ T. At the intermediate temperatures, i.e., $T_N/2 < T < T_N$, the MR exhibits stepwise decrease with increasing field except for a slight initial increase at low fields. Such a transition from negative to positive MR is also found in the $R(T)$ curves for $H = 9$ T and 0 T in Fig.1a, where a crossing at $T \approx T_N/2$ is clearly identifiable (see more $R(T)$ curves of Sample A and Sample B in Fig.S2). As presented in the magnetic phase diagram in Fig.S1b, the steps in the $R(H)$ curves at $T_N/2 < T < T_N$ correlate with the magnetic phase transitions determined by magnetization measurements. On the other hand, the $R(H)$ curves at $T < T_N/2$ show no clear steps, although magnetization measurements also reveal various magnetic phase transitions in this temperature regime (Fig.S1).

Figure 2a shows $R(\theta)$ curves obtained at a constant rotating magnetic field ($H_R = 6$ T) and at various temperatures and Fig.2b shows those measured at a fixed temperature ($T = 20$ K) and various $H_R$. Fig.S3 presents results for Sample A obtained at $T = 3$ K and in various magnetic fields and Fig.S4 shows results for Sample B. MR anisotropy can be clearly identified in both Sample A and Sample B. As presented below, the $R(\theta)$ behavior enables us to uncover the interplay of magnetism and electro-transport in CeBi.

CeBi has a rock-salt cubic crystal lattice, resulting in identical electronic structures along the $k_x$, $k_y$ and $k_z$ directions of the Brillouin zone [47]. Similar to its non-magnetic rare-earth monopnictide counterpart LaSb [49], CeBi has a bulk anisotropic Fermi surface with elongated electron Fermi pockets [38,47]. Thus, we expect to see anisotropic MR in CeBi in a rotating magnetic field. Indeed, some of the $R(\theta)$ curves in Fig.2 resemble those found in LaSb [49], showing a four-fold symmetry with maxima at $\mathbf{H} \parallel <011>$, e.g., curves for $H_R = 6$ T and $T > 20$ K in Fig.2a as well as for 20 K and $H_R > 6$ T in Fig.2b. However, unlike those in LaSb [49], the MR anisotropy in CeBi varies strongly with both the temperature and the magnitude of the magnetic field. For example, the peaks in the $R(\theta)$ curves at $\mathbf{H} \parallel <011>$ in Fig.2a for $H_R = 6$ T at $T > T_N/2$ transform into dips at $T < T_N/2$. Fig.2b also shows that the shape of the $R(\theta)$ curves evolves drastically between $H_R = 2$ T and 8 T and especially becomes most complex at 3.3 T $< H_R <$ 4.0 T. The values of the MR for $\mathbf{H} \parallel <011>$ are indeed higher than those for $\mathbf{H} \parallel \mathbf{b}$ and $\mathbf{c}$. Furthermore, the MRs for $\theta = 0°$ and $90°$ are also different, most significantly at 2.2 T $< H_R <$ 2.8 T (Fig.S5), probably due to the temperature-induced tetragonal distortion at $T < T_N$, which breaks the cubic lattice symmetry [43].

Recently, Xu et al. [50] investigated the MR anisotropy of CeSb, which is a sister cerium monopnictide of CeBi, in the low temperature high magnetic-field driven ferromagnetic state. Minima in $R(\theta)$ curves at $\mathbf{H} \parallel <011>$ were also observed and attributed to the flops of the magnetization dominated by the planar $\Gamma_8$ orbitals of the Ce $f$ electrons [51,52]. Due to strong $p$-$f$ mixing, CeSb has a strong magnetic anisotropy with the easy axis along $<001>$ [46]. The magnetization is induced only by the component of the magnetic field along $<001>$, resulting in a $1/cos\varphi$ dependence of the magnetic field values $H_{FM}$ above which the material is in the ferromagnetic phase [50,53], where the relationship of $\varphi$ to the angle $\theta$ defined in the inset of Fig.2 is $\varphi = \theta - n\pi/2$ with $n = 0, 1, 2, 3$, and 4 for $0° \leq \theta \leq 45°$, $45° \leq \theta \leq 135°$, $135° \leq \theta \leq 225°$, $225° \leq \theta \leq 315°$, and $315° \leq \theta \leq 360°$, respectively [50]. In Fig.S6a we present the $R(H)$ curve obtained at $T = 3$ K and $\theta = 45°$, from which $H_{FM}$ can be derived. Fig.S6b presents the angle dependence of $H_{FM}$. It clearly shows that $H_{FM}$ indeed follows a $1/cos\varphi$ dependence. That is, the MR and its anisotropy in CeSb and CeBi at low temperatures, particularly in high magnetic fields may have similar origins. For example, the dips in $R(\theta)$ curves presented in Fig.S3 for **H** ∥ <011> and $T = 3$ K originate most likely from magnetization-flops, similar to those in CeSb [50]. However, Fig.S7 also discloses significant differences of CeBi's MR behavior from that of CeSb. For example, at the ferromagnetic transition, CeSb has lower resistances in the ferromagnetic phase for all field orientations [50,51]. The $R(H)$ curves of CeBi show only a slight downturn kink at $\theta = 0°$ or an upturn bump at $\theta = 90°$ (insets in Fig.S7). At other angles the resistances of CeBi in the ferromagnetic phase become higher.

Similar to that in CeSb [50], we are unable to offer a quantitative analysis on the anisotropy of the positive MR in CeBi (at $T < T_N/2$), as demonstrated by the analysis on $R(\theta)$ at $T = 3$ K in Fig.S8. However, magnetization-flops revealed for the ferromagnetic phase at low temperatures ($T < T_N/2$) may also exist at high temperatures ($T > T_N/2$) and in non-ferromagnetic magnetic phases. For this purpose, we measured $R(H)$ curves at various angles at $T = 20$ K, with data at $0° \leq \theta \leq 45°$ presented in Fig.3a. Similar to that presented in Fig.1b for **H** ∥ **c**, three magnetic phase transitions can be clearly identified in $R(H)$ curves for all field orientations. Furthermore, all curves look very similar, except those for $43° \leq \theta \leq 45°$, i.e., ~2° within the <011> directions, which have a sharper initial step. More importantly, the $R(H)$ curve shifts to higher field values when the magnetic field is rotated towards [011] ($\theta = 45°$) from the **c**-axis. As presented in Fig.S9, the angle dependence of the magnetic field $H_i$ (i = 1,2,3) at which a magnetic phase transition occurs indeed exhibits the same hallmark of magnetization-flop as the $H_{FM}$ in Fig.S6b for $T = 3$ K, i.e., following a $H_i \sim 1/cos\varphi$ relationship for all three transitions. This indicates that when the magnetic field is rotated across the <011> direction, the magnetization **M** that orients along one of the primary crystalline axes flops to another primary crystalline axis that is within 45° to the magnetic field, akin to that in the ferromagnetic phase at $T < T_N/2$. It further reveals that the magnetic phase transition is induced by the component $\mathbf{H_M}$ of the magnetic field $\mathbf{H_R}$ along **M**, with the value of $H_M = H_R cos\varphi$ and $H_R$ being the amplitude of the rotating magnetic field. This also implies that when the magnetic field tilts away from a primary crystalline axis it requires a higher magnitude to generate the same magnetic state in the material. In fact, as presented in Fig.3b, the $R(H)$ curves obtained at $\varphi \neq 0°$ nearly overlap with that for $\varphi = 0°$ if the magnetic field values are scaled as $Hcos\varphi$. The slight difference probably comes from the demagnetization effect and the small misalignment of the magnetization to the primary crystalline axis, as theoretically revealed for the ferromagnetic state in CeSb [50].

The above scaling behavior for $R(H)$ curves at various angles indicates the angular dependence of the resistance at a constant $H_R$ is due to the change of the effective field $H_R cos\varphi$ responsible for producing the magnetic state. Thus, we can use the $R(H)$ curve obtained along a primary crystalline axis to derive the $R(\theta)$ curve at a constant $H_R$. For example, the resistance value at a field $H$ of the $R(H)$ curve at $\varphi = 0°$ should be the same as that in the $R(\theta)$ curve at $\theta = n\pi/2-\varphi$ with $\varphi = arccos(H/H_R)$ obtained at a fixed rotating field $H_R$. As discussed above, the $R(H)$ curves at $\theta = 0°$ and 90° are not the same, although $\varphi = 0°$ in both cases. In the conversion we used the $R(H)$ curve at $\theta = 90°$ (**H ∥ b**) to construct the $R(\theta)$ section at $45° \leq \theta \leq 135°$ in which **M** remains along the

**b**-axis. Similarly, the $R(H)$ curve at $\theta = 0°$ (**H** ∥ **c**) was used to calculate the $R(\theta)$ section at $0° \leq \theta \leq 45°$ and $135° \leq \theta \leq 180°$ (see Fig.S10 for more conversion details). As presented in Fig.4 for representative $R(\theta)$ curves at $T = 20$ K, the calculated and experimental results are coincident, with the small deviations due to the possible causes cited in the preceding paragraph.

The scaling in Fig.3b as well as the consistency of the calculated and experimental $R(\theta)$ curves in Fig.4 indicate that the MRs for all magnetic field orientations are solely governed by the magnetization which aligns along a primary crystalline axis and flops in a rotating magnetic field. The disparity between the $R(H)$ curves obtained for $\theta = 0°$ and $90°$ (Fig.S5), the reflection symmetry to the **b**-axis in the $R(\theta)$ curves at $45° \leq \theta \leq 135°$ and to the **c**-axis in the $R(\theta)$ curves at $0° \leq \theta \leq 45°$ and $135° \leq \theta \leq 180°$ in Fig.4 provide additional evidence of the direct effect of the magnetization on the MR and its anisotropy. However, these remarkable behaviors are exclusive to $T_N/2 < T < T_N$. As presented in Fig.S8 for the $R(\theta)$ curve at $T = 3$ K, the MR anisotropy for $T < T_N/2$ cannot be derived from $R(H)$ curves measured at $\theta = 0°$ and $90°$. A molecular field model [54] assuming only $p$-$f$ mixing can account for the experimental magnetic transitions in CeBi. It reveals that the molecular field coefficient is a sensitive function of temperature at $T_N/2 < T < T_N$ and remains unchanged at $T < T_N/2$. Optical conductivity [55] and ARPES [56] experiments also detect transitions from $p$-$f$ mixing to $p$-$f + p$-$d$ mixings and from a double Dirac-cone band to a single Dirac-cone band at $T \approx T_N/2$ when the temperature is lowered. They are consistent with the change of the MR anisotropy revealed here.

The uniqueness of the observed magnetoresistance anisotropy in CeBi at $T_N/2 < T < T_N$ is further corroborated in its topologically trivial counterpart CeSb [57,58]. Similar to CeBi which is in the ground state Type IA at low temperatures ($T < T_N/2$), CeSb also has a transition to its ground state Type IA at $T \approx 8$ K ($\sim T_N/3$) upon cooling in zero field [50]. They both show similar MR anisotropy at low temperatures (see Fig.S3 and related discussion). As presented in Fig.S11, however, $R(H)$ curves obtained for CeSb at $T > T_N/3$, e.g., $T = 11$ K, can be non-monotonic, in contrast to the pure negative MR in CeBi at $T > T_N/2$. More importantly, the $R(H)$ curves for various angles do not follow the scaling behavior presented in Fig.3, although the deviation becomes smaller when the $R(H)$ curves show pure negative MR, e.g., for $T = 19$ K. Figure S12 presents a comparison of the experimental $R(\theta)$ curves with those calculated by assuming a magnetization-governed MR anisotropy for $T = 19$ K. Clearly, the discrepancy is more pronounced than that in Fig.4 for CeBi.

In summary, we investigated the magnetoresistance behavior of the magnetic topological semimetal CeBi in a rotating magnetic field. We identified a temperature regime ($T_N/2 < T < T_N$) in which the magnetoresistance anisotropy is purely controlled by the magnetization that orients along a primary crystalline axis and flops under the influence of a rotating magnetic field, revealing unusually strong interaction between conduction electrons and magnetization. The results demonstrate that CeBi can be a versatile platform for exploring the interplay of magnetism and eletro-transport as well as orbital physics in magnetic topological materials.

**Acknowledgment**

Magnetotransport measurements were supported by the U.S. Department of Energy, Office of Science, Basic Energy Sciences, Materials Sciences and Engineering. Magnetization characterizations were supported by the National Science Foundation under Grant No. DMR-1901843. Y. Y. L, Y. L. W and H. B. W. acknowledge supports by the National Natural Science Foundation of China (61771235 and 61727805). F. H. and M. L acknowledge support from U.S. Department of Energy (DOE), Office of Science, Basic Energy Science (BES) award No. DE-SC0020148.

**Figure captions**

**Fig.1**. (a) Temperature dependence of the resistance at $H = 0$ T and 9 T. (b) Magnetic field dependence of the resistance at various temperatures. The data are from Sample A and the magnetic field is parallel to the **c**-axis. Upon cooling in zero field, the transition at $T_N$ is from paramagnetic to antiferromagnetic (AFM) and at $T_N/2$ is from Type I (↑↓↑↓) AFM to Type IA (↑↑↓↓) AFM, with ↑ and ↓ representing magnetic moment orientations. In (b), stepwise changes in the magnetoresistance occur at $T_N/2 < T < T_N$. The dashed and dotted lines are a guide to the eye for $H = 0$ T and 9 T, respectively.

**Fig.2**. Angle dependence of the magnetoresistance of Sample A in $H_R = 6$ T and at various temperatures (a) and at $T = 20$ K and in various magnetic fields (b). The dotted lines are a guide to the eye for the magnetoresistance at $\theta = 45°$ and 135°. The inset is a schematic showing the definition of the angle $\theta$ for the magnetic field orientation.

**Fig.3**. (a) Magnetic field dependence of the resistance of Sample A at $T = 20$ K and at various angles. (b) Replot of the data in (a), with the applied field value $H$ replaced with the component $H\cos\varphi$ along the primary crystalline axis, where $\varphi$ is the angle between **H** and the closest primary crystalline axis.

**Fig.4**. Comparison of the experimental $R(\theta)$ data (symbols) of Sample A at $T = 20$ K for various rotating magnetic fields $H_R$ and those derived (lines) from the $R(H)$ curves obtained with **H** along the **b**- and **c**-axis, by converting $H$ into $\varphi$ using $\varphi = \arccos(H/H_R)$ and $\theta = n\pi/2 - \varphi$ (see Fig.S10 for more details on the analysis). The value of the $H_R$ is indicated for each curve in the figure.

**Figure 1**

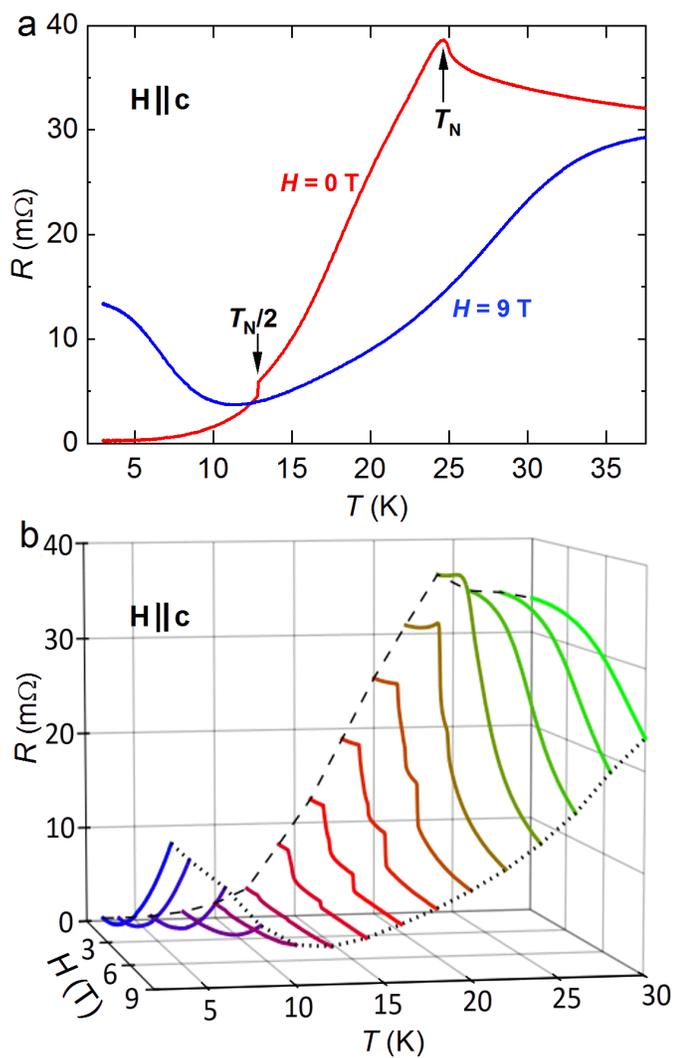

**Figure 2**

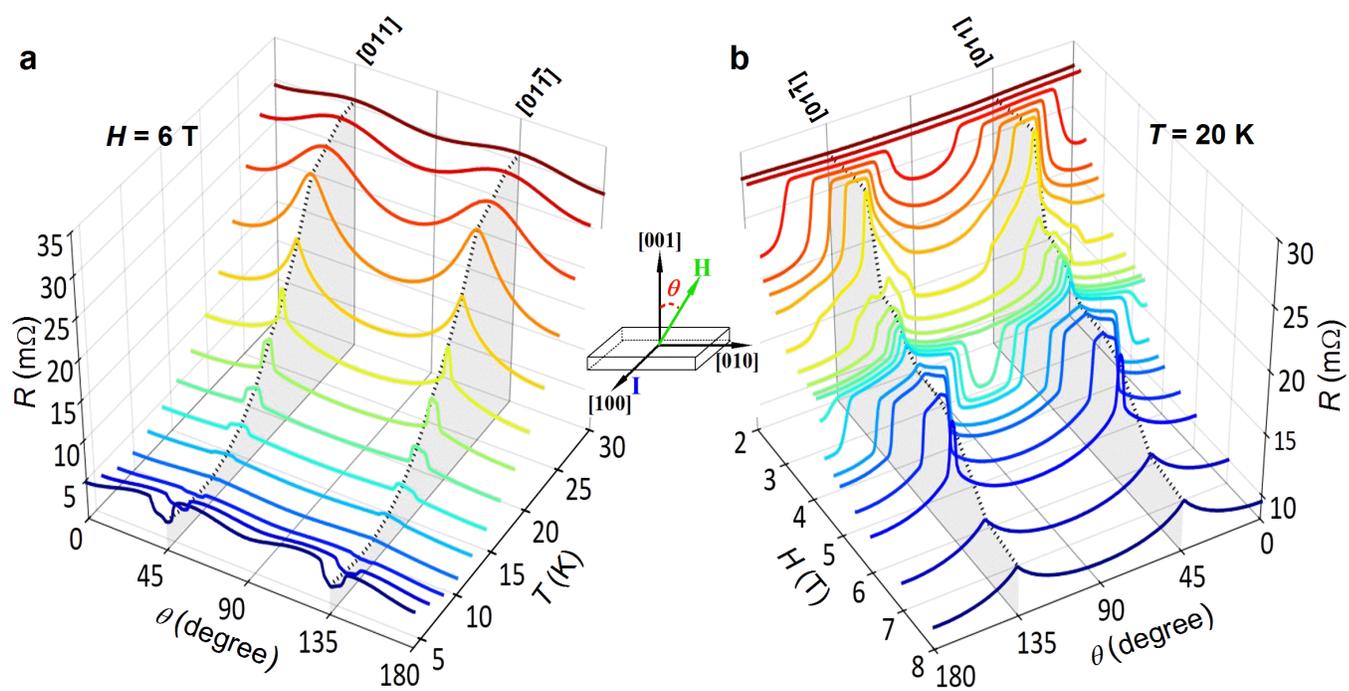

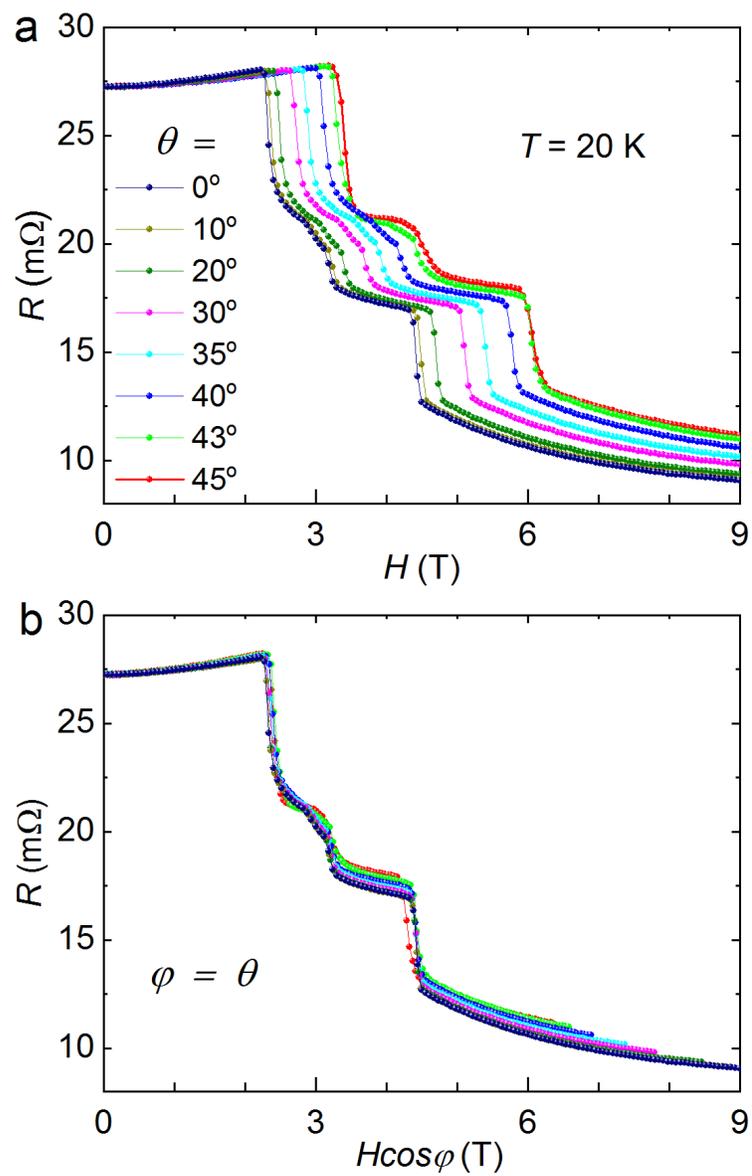

**Figure 4**

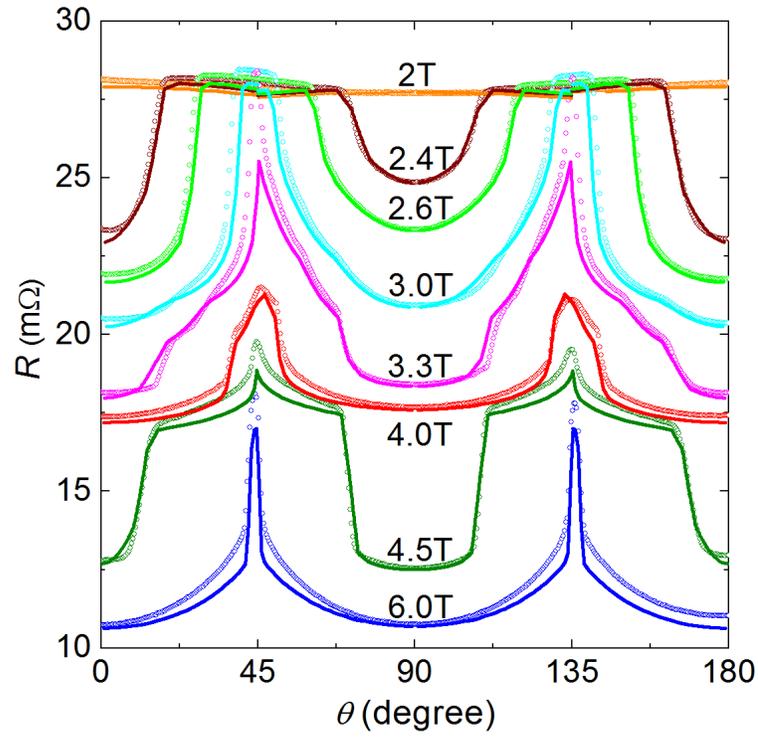

**Magnetization Governed Magnetoresistance Anisotropy in Topological Semimetal CeBi**

by Yang-Yang Lyu et al.

**Crystal growth:** CeBi single crystals were grown with a Bi self-flux method. The Ce-Bi binary phase diagram reported in Ref.1 reveals the existence of a quasi-triangle "CeBi + liquid" regime for Ce concentration of 18% to 50% (atomic percentage), enabling us to grow CeBi crystals using Bi as the flux. We chose a Ce atomic percentage of 23%, corresponding to a temperature range of ~1100 °C to ~910 °C for the CeBi crystal growth (below 910 °C CeBi$_2$ will precipitate instead). In experiments, Ce ingots (99.9%, Alfa Aesar) and Bi ingots (99.999%, Beantown Chemicals) with an atomic ratio of Ce:Bi = 23:77 were loaded into an alumina crucible, which was covered with a stainless steel frit and sealed in a quartz tube of 16 mm in diameter under a vacuum of $10^{-3}$ mbar. The quartz tube was heated up to 1150 °C in 10 hours, kept at that temperature for another 10 hours, then slowly cooled down to 910 °C at a rate of 3 °C/h. Finally, the quartz tube was centrifuged at 910 °C to separate CeBi crystals from Bi flux. Large polyhedral CeBi crystals with dimensions up to 5×5×4 mm$^3$ and minor amount of plate-like CeBi$_2$ crystals were harvested. CeBi and CeBi$_2$ crystals can be distinguished easily by their morphologies. CeBi crystals were characterized by single crystal X-ray diffraction on the diffractometer STOE IPDS 2T and a rock-salt type structure was confirmed. Single crystal X-ray diffraction experiments revealed the straight edges of the cleaved CeBi crystals to be the *a/b/c* axis. Information on CeSb crystal growth can be found in Ref.2.

**Resistance measurements:** We conducted DC resistance measurements on two CeBi crystals (Sample A and Sample B) in a homemade magnetotransport system and one CeSb crystal in a Quantum Design PPMS-9 using constant current mode ($I$ = 1 mA) in a magnetic field up to 9 Tesla. The electric contacts were made by attaching 50 μm diameter gold wires using silver epoxy, followed with baking at 120 °C for 20 minutes. Angle dependence of the resistance was obtained by placing the sample on a precision, stepper-controlled rotator (Attocube model ANR 51 for the homemade system and HR 4084-304 for PPMS-9) with an angular resolution of 0.05°. The inset of Figure 2 shows the measurement geometry where the magnetic field **H**($\theta$) is rotated in the (100) plane and the current **I** flows along the [100] direction, such that the magnetic field is always perpendicular to the applied current **I**. We define the magnetoresistance as $MR = [R - R_0)]/R_0$, where $R$ and $R_0$ are resistivities at a fixed temperature with and without magnetic field, respectively.

**Magnetization measurements:** We measured the magnetic field dependence of the magnetization of a CeBi crystal (Sample C) at various temperatures using a Quantum Design MPMS XL. The magnetic field was aligned along a primary crystalline axis.

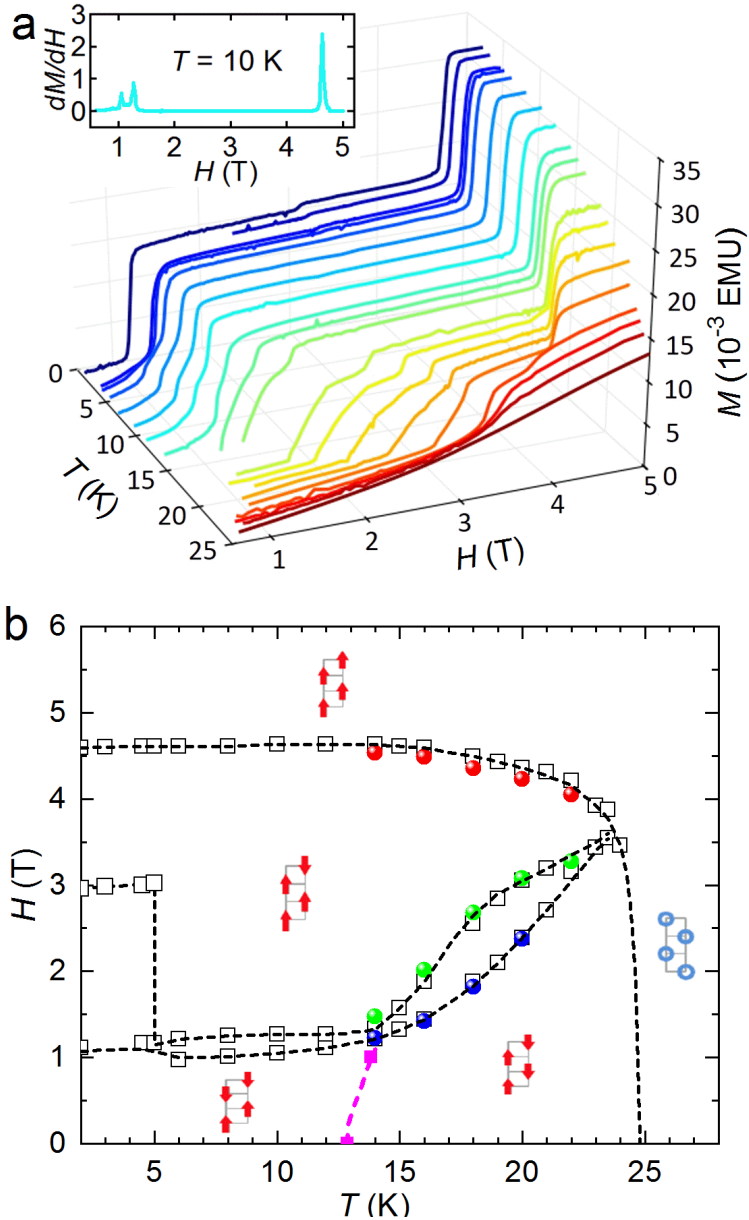

**Fig.S1**. (a) Magnetic field dependence of the magnetization $M(H)$ of Sample C at **H‖c** and at various temperatures. The inset shows a representative $dM/dH$ versus $H$ curve, through which the transition fields are determined from the peak field values. (b) Experimentally determined magnetic phase diagram. Open symbols represent results derived from $M(H)$ data in (a). Solid circles and squares are results obtained from $R(H)$ curves in Fig.1b and $R(T)$ curves in Fig.1a and Fig.S2, respectively. Dashed lines are a guide to the eye. The schematics represent the magnetic structures of the Ce layers, where red arrows indicate the directions of the magnetic moments in the ordered Ce-ion layers with magnetic moments of $2\mu_B$/Ce and blue circles indicate paramagnetic ones. In zero field, the anti-ferromagnetic phases are type I and type IA above and below $T_N/2$ ($\approx 12.8$ K), respectively. The phases without schematics have more complicated magnetic structures [Please refer to Ref.3].

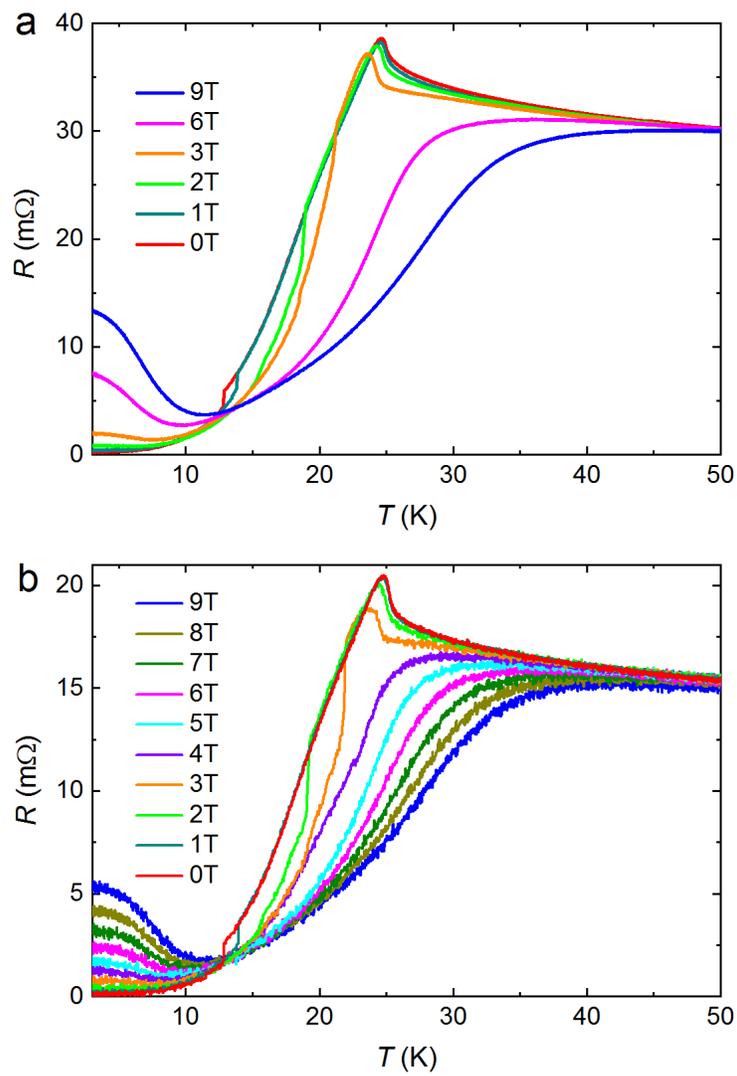

**Fig.S2**. Temperature dependence of the resistance at various magnetic fields for Sample A (a) and Sample B (b).

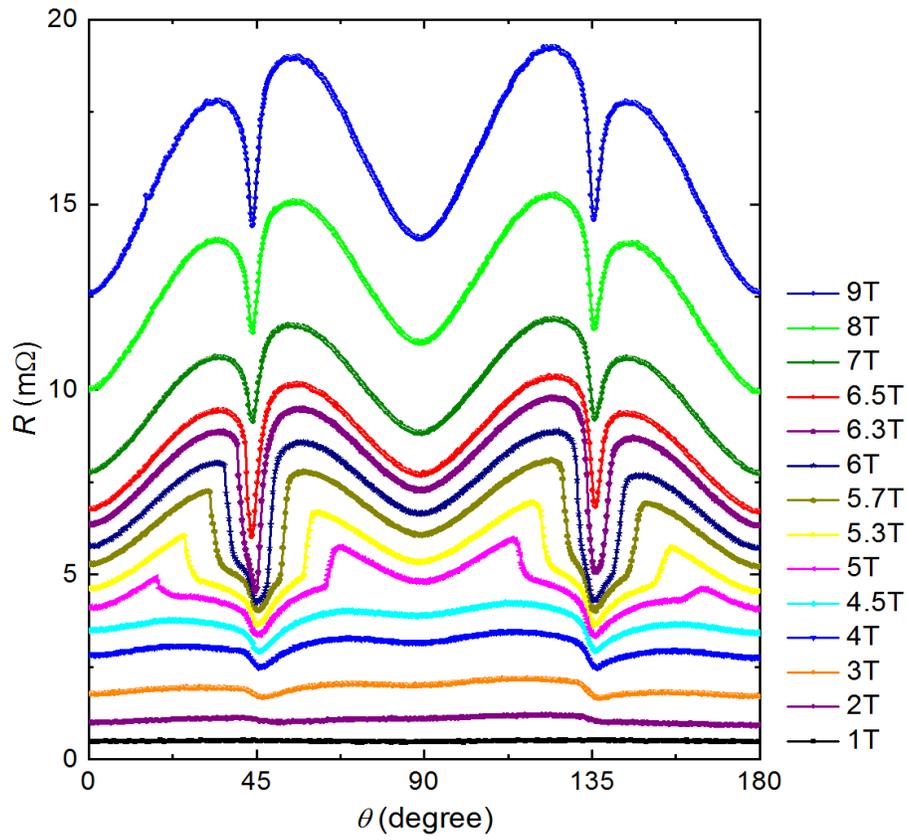

**Fig.S3**. Angle dependence of the resistance of Sample A at $T = 3$ K and under various magnetic field values, given in the right side of the figure.

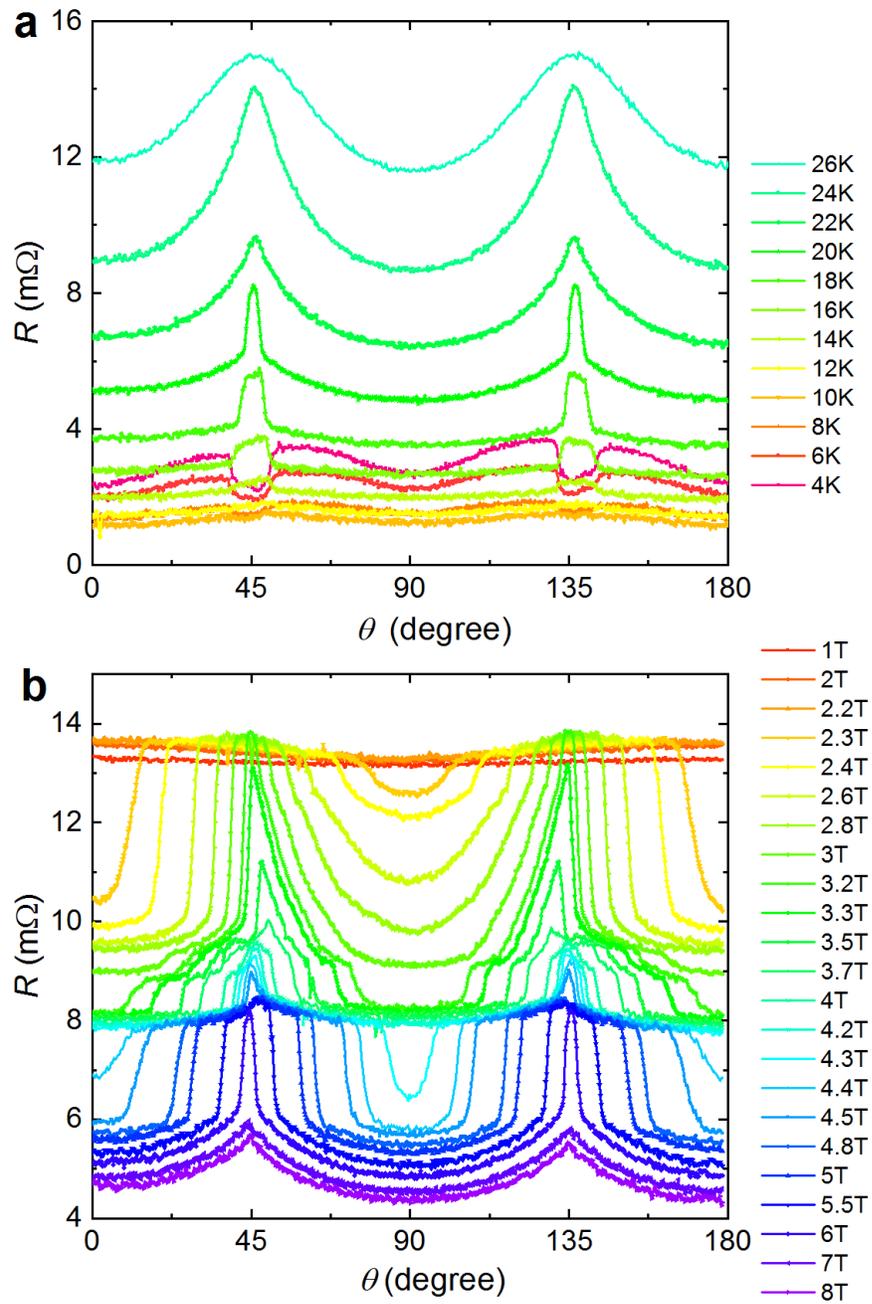

**Fig.S4**. Angle dependence of the magnetoresistance of Sample B. (a) at $H = 6$ T and various temperatures. (b) at $T = 20$ K and various magnetic fields.

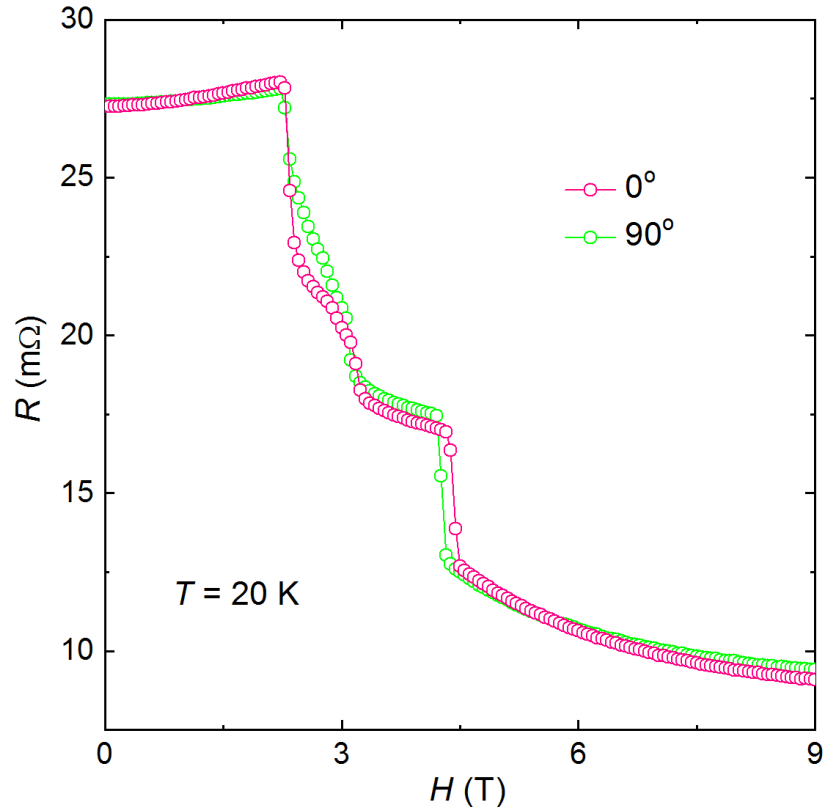

**Fig.S5**. Magnetic field dependence of the resistance of Sample A at $\theta = 0°$ and $90°$. The magnetoresistance behavior at **H ‖ b** ($\theta = 90°$) and **H ‖ c** ($\theta = 0°$) is not exactly the same as that expected for a cubic crystal lattice, probably due to a temperature induced tetragonal distortion (see text for the associated reference).

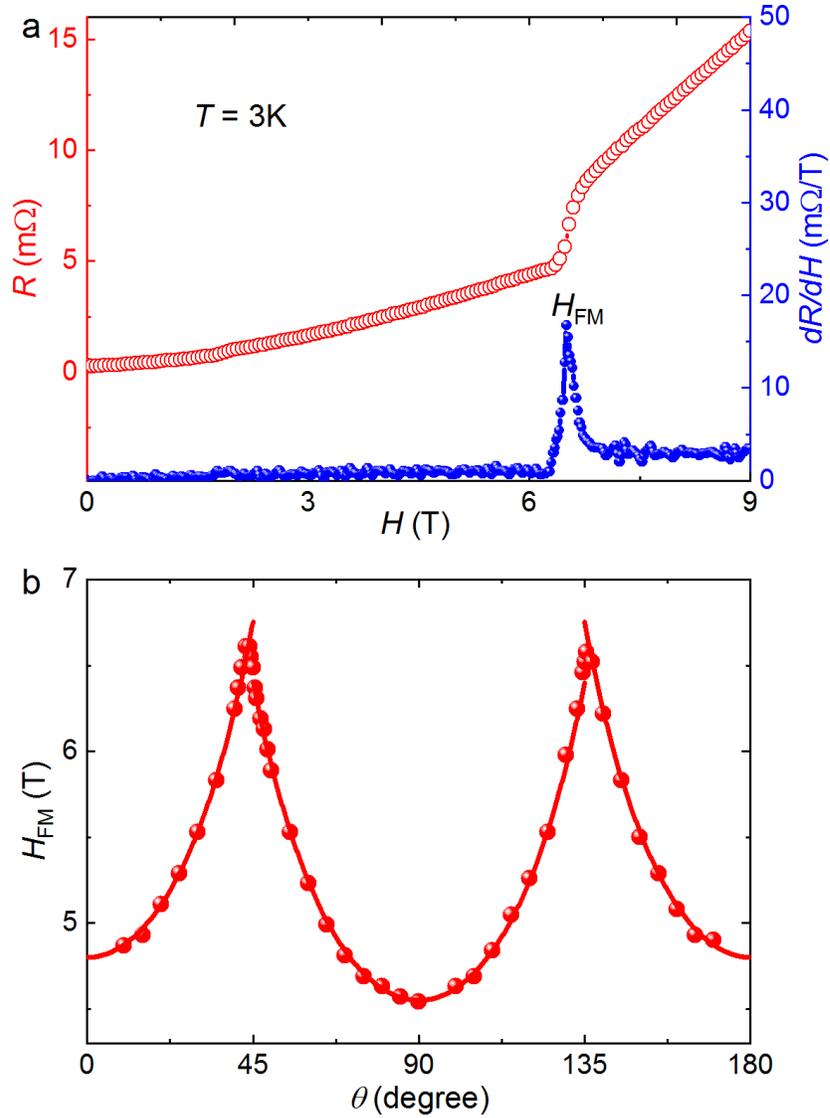

**Fig.S6**. (a) $R(H)$ curve of Sample A at $\theta = 45°$ and its derivative $dR/dH$ (more $R(H)$ curves are presented in Fig.S7). The ferromagnetic transition field $H_{FM}$ is determined from the peak in the $dR/dH(H)$ curve. (b) Angle dependence of the magnetic transition field. Symbols are experimental results and lines are fits using $H_{FM} = H_{FM0}/cos\,\varphi$ with $\varphi = \theta - n\pi/2$ with $n = 0, 1, 2$ for $0° \leq \theta \leq 45°$, $45° \leq \theta \leq 135°$, $135° \leq \theta \leq 180°$, respectively (see text for details), with $H_{FM0} = 4.80$ T for $0° \leq \theta \leq 45°$ and $135° \leq \theta \leq 180°$; $H_{FM0} = 4.55$ T for $45° \leq \theta \leq 135°$.

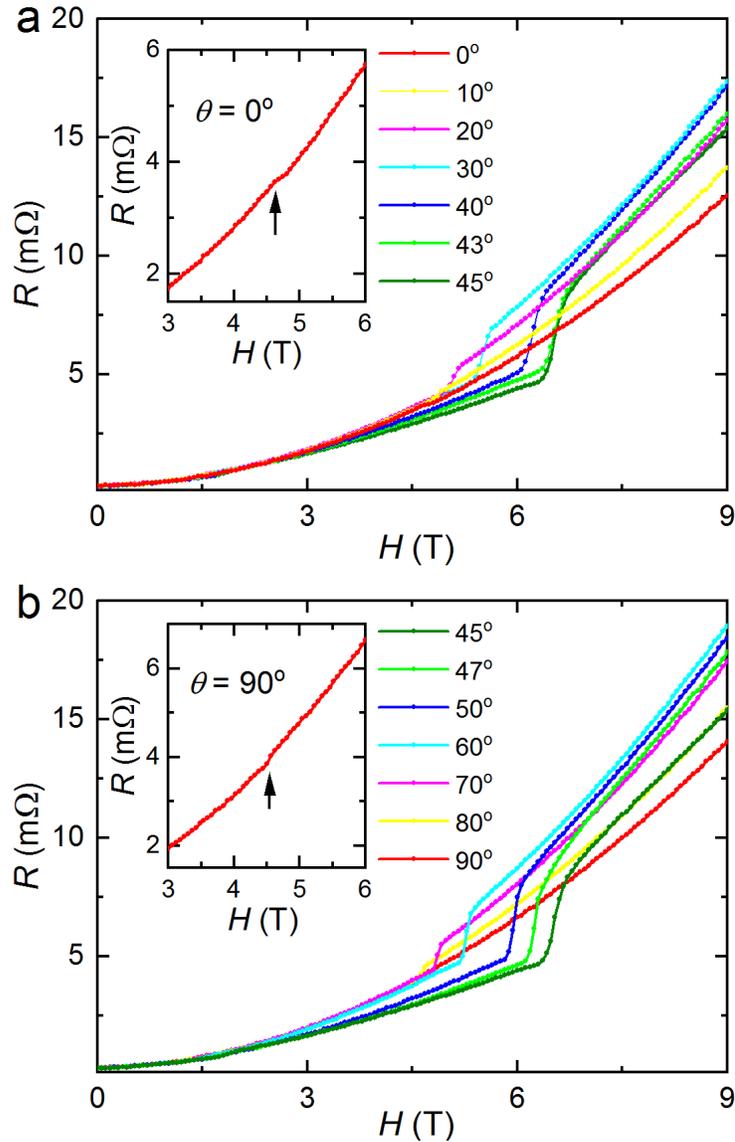

**Fig.S7**. $R(H)$ curves of Sample A at $T = 3$ K. (a) at $\theta = 0° - 45°$; (b) at $\theta = 45° - 90°$. They show clear steps for magnetic fields orientated away from the primary axis ($\theta = 0°$ and $90°$). The insets present expanded views near the expected ferromagnetic (FM) transition for the $R(H)$ curves measured at $\theta = 0°$ and $90°$. Arrows indicate the locations of a small kink, likely originating from the FM transition.

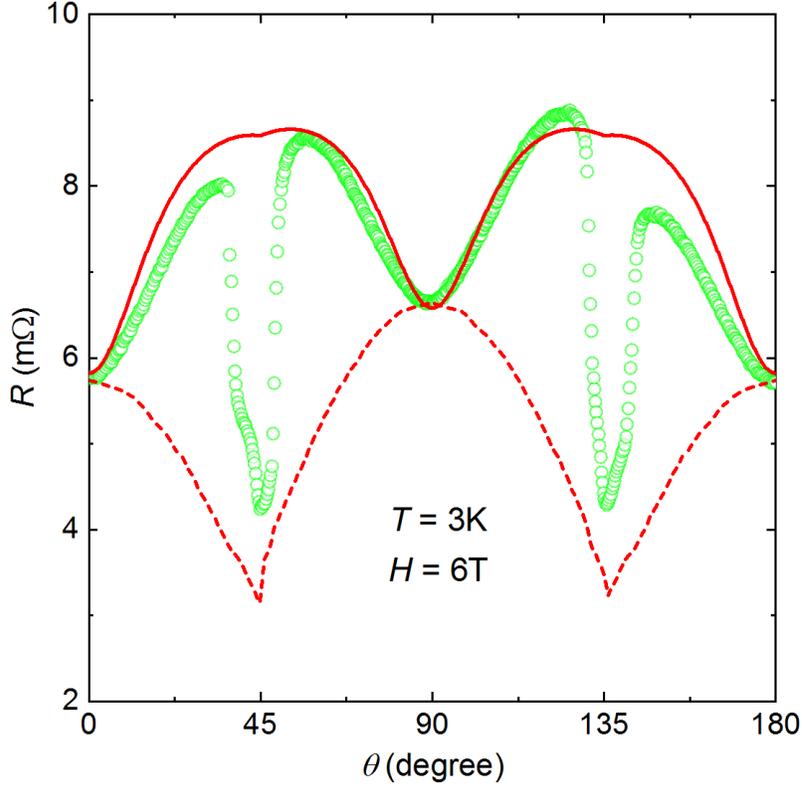

**Fig.S8**. Comparison of the experimental $R(\theta)$ data (symbols) of Sample A at $T = 3$ K for a rotating magnetic field $H_R = 6$ T and those (dashed lines) derived from the $R(H)$ curves (see Fig.S7) obtained with **H** along **b**- and **c**-axis ($\theta = 0°$ and $90°$), by converting $H$ into $\varphi$ using $\varphi$ = arccos($H/H_R$), where $|\varphi| \leq 45°$ is the angle between the magnetic field and the primary axis and $\theta = n\pi/2$- $\varphi$ (see text for more details). The solid curve represents the calculated results expected for an anisotropic Fermi surface with anisotropic induction $B = [H^2 + M_0^2 + 2HM_0 cos\varphi]^{1/2}$. $M_0 = 0.38$ T was used for a ferromagnetic CeBi with a Ce magnetization of 2 $\mu_B$/atom, where $\mu_B$ is the Bohr magneton, using the analysis procedures in Ref.2. We used two elliptical electron pockets with an anisotropy of $\lambda_\mu = 4$ and one isotropic hole pocket. The derived mobilities are $\mu_{//} = 0.55$ and 0.83 $m^2V^{-1}s^{-1}$, $\mu_\perp = 8.8$ and 13.2 $m^2V^{-1}s^{-1}$ for electrons, and $\mu_h = 0.85$ $m^2V^{-1}s^{-1}$ for holes. The different electron mobilities in the two perpendicular pockets are required to account for the asymmetry in the experimental $R(\theta)$ curve, which could originate from the lattice distortion as discussed in the text. The derived mobilities are very close to those of CeSb [2].

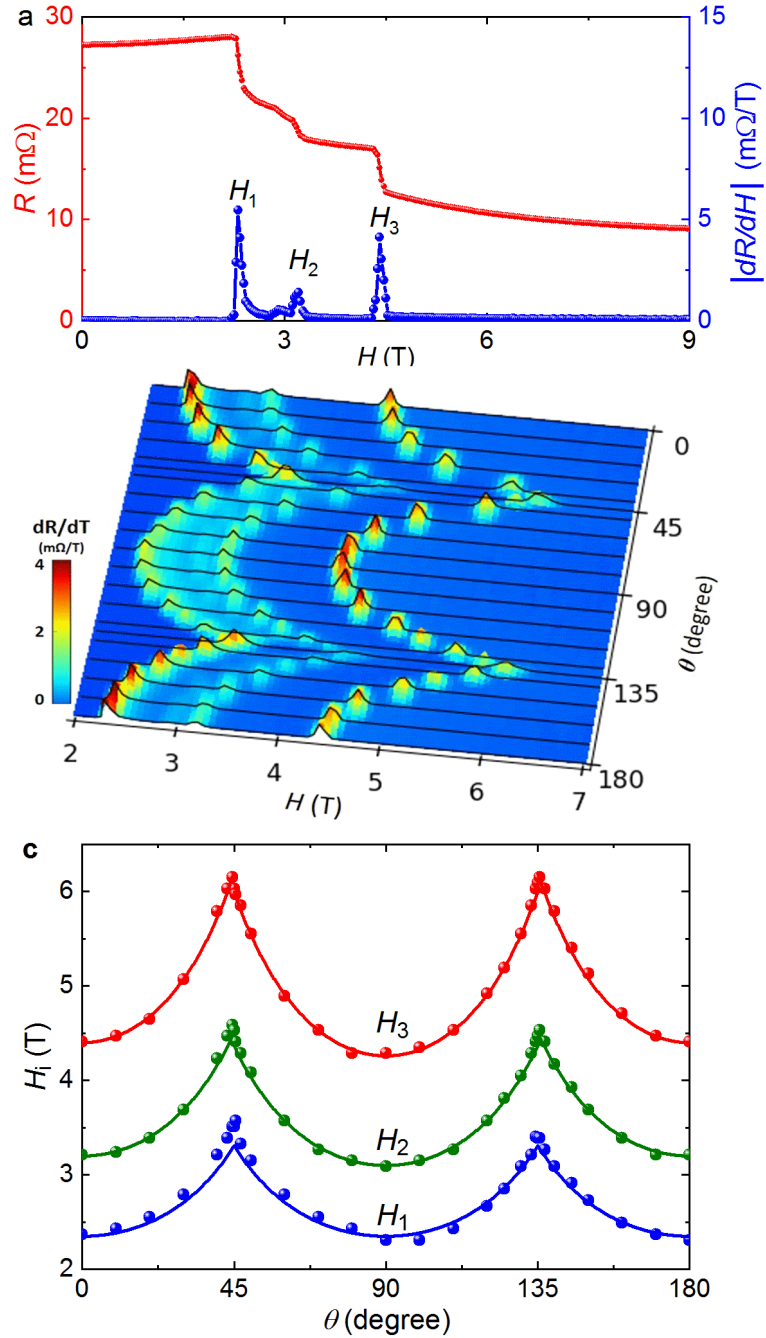

**Fig.S9**. (a) $R(H)$ curves of Sample A at $\theta = 0°$ and its derivative $dR/dH$. Three transition fields $H_i$ (i = 1,2,3) are determined from the peaks in the $dM/dH$ versus $H$ curve. (b) Color map of $dR/dH(H)$ for all measured angles. (c) Angle dependence of the magnetic transition fields. Symbols are experimental and lines are fits using $H_i = H_{i0}/cos\varphi$ with $\varphi = \theta - n\pi/2$ with $n = 0, 1, 2$ for $0° \leq \theta \leq 45°$, $45° \leq \theta \leq 135°$, $135° \leq \theta \leq 180°$, respectively (see text for details). For $0° \leq \theta \leq 45°$ and $135° \leq \theta \leq 180°$, $H_{10}$, $H_{20}$ and $H_{30}$ are 2.35 T, 3.2 T and 4.4 T, respectively; For $45° \leq \theta \leq 135°$, $H_{10}$, $H_{20}$ and $H_{30}$ are 2.35 T, 3.1 T and 4.26 T, respectively. All data were taken at $T = 20$ K.

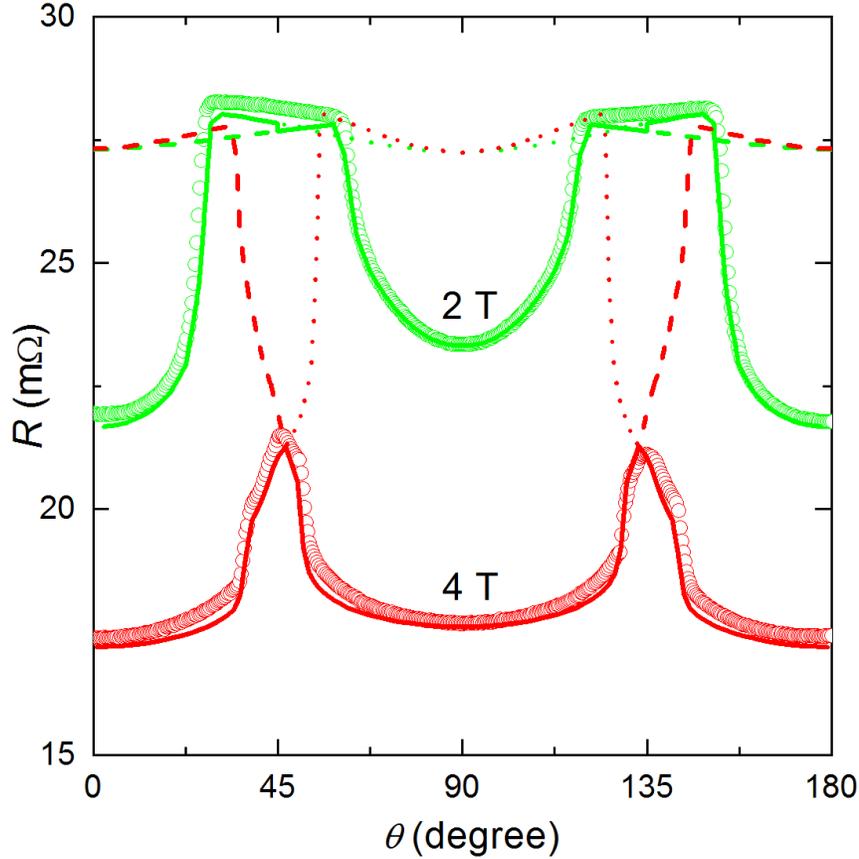

**Fig.S10**. Comparison of the experimental data (symbols) for $H_R = 2$ T and 4 T with the $R(\theta)$ curves (solid lines) calculated from the $R(H)$ curves measured at $\theta = 0°$ and 90°, where $H$ at $\theta = 0°$ and 90° is first converted into $\varphi$ with $\varphi = arccos(H/H_R)$ for a fixed rotating field $H_R$ and then into $\theta$ using $\theta = n\pi/2 - \varphi$. Obviously, only the section of the $R(H)$ curve with $H \leq H_R$ can be converted into the $R(\theta)$ curve. Since the $R(H)$ curves at $\theta = 0°$ and 90° are not the same, the section of the $R(\theta)$ curve at $0° \leq \theta \leq 45°$ and $135° \leq \theta \leq 180°$ is calculated from the $R(H)$ curve measured at $\theta = 0°$ while that at $45° \leq \theta \leq 135°$ is converted from the $R(H)$ curve measured at $\theta = 90°$. Dotted and dashed lines are the sections of $R(\theta)$ curves calculated also from the $R(H)$ curve measured at $\theta = 0°$ and $\theta = 90°$, respectively. They are not experimentally detectable due to magnetization (orbital)-flop when **H** is rotated across $\theta = 45°$ and 135°.

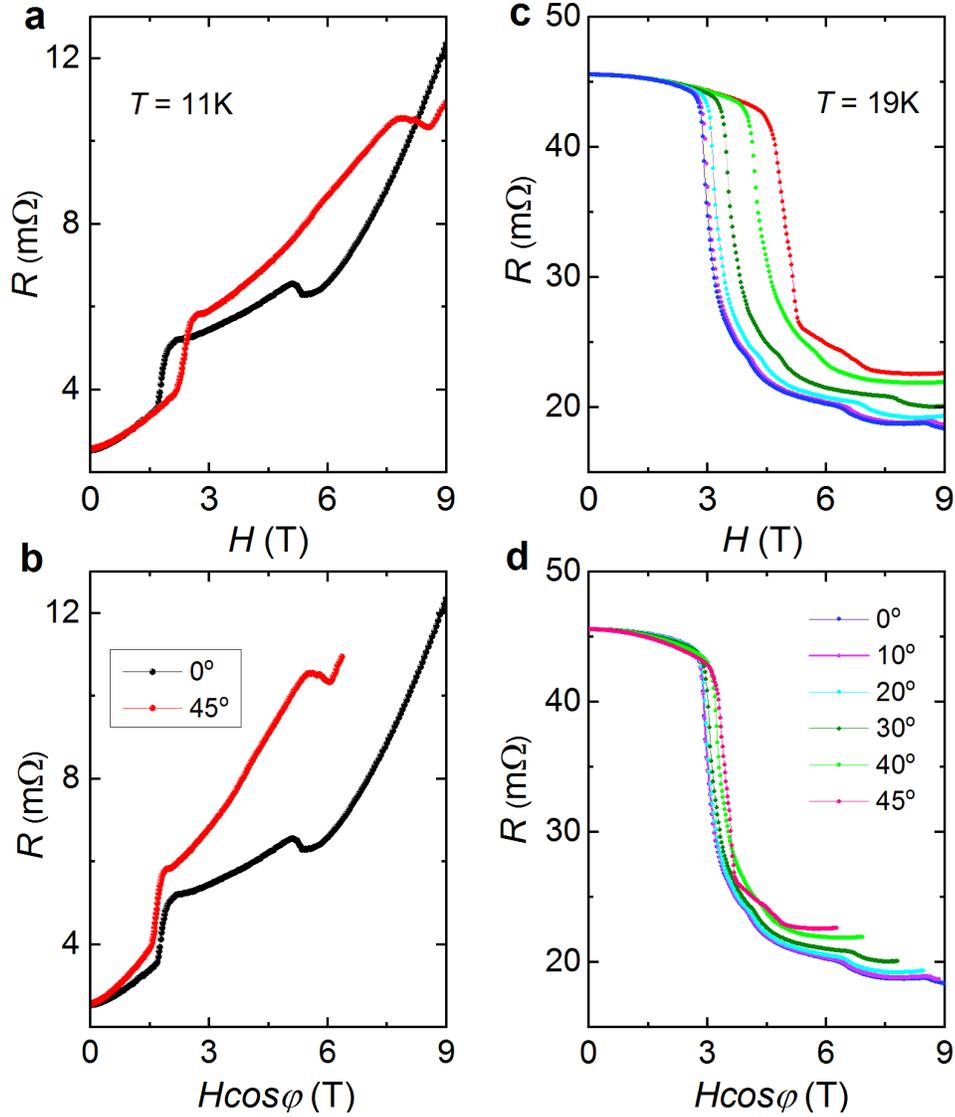

**Fig. S11**. Absence of scaling behavior for the anisotropic magnetoresistance in CeSb. (a, c) Magnetic field dependence of the resistance of a CeSb crystal at $T = 11$ K and 19 K, respectively. (b, d) Corresponding replots of the data in (a, c), with the applied field value $H$ replaced with the field component $H\cos\varphi$ along the primary crystalline axis, where $\varphi$ is the angle between **H** and the closest primary crystalline axis.

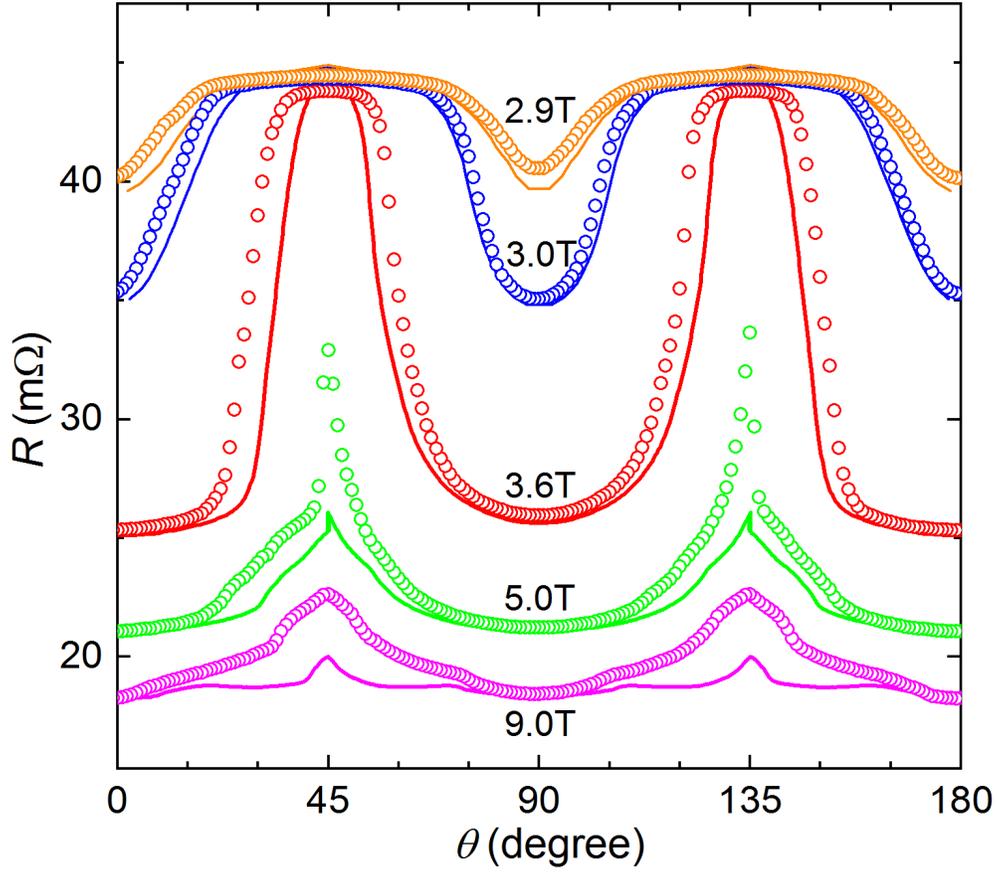

**Fig.S12**. Comparison of the experimental and the calculated $R(\theta)$ curves for a CeSb crystal. Symbols are experimental data at $T = 19$ K for various magnetic fields $H_R$. The solid lines are calculated from the $R(H)$ curves measured at $\theta = 0°$ and $90°$, where $H$ at $\theta = 0°$ and $90°$ is first converted into $\varphi$ with $\varphi = arccos(H/H_R)$ for a fixed rotating field $H_R$ and then into $\theta$ using $\theta = n\pi/2 - \varphi$ (see Fig.S10 for more details on the analysis). The value of the $H_R$ is indicated for each curve in the figure.

**Supplementary References**